\documentclass{article}

\usepackage[numbers]{natbib}         
\usepackage[colorlinks]{hyperref}    
\usepackage[english]{babel} 
\usepackage{amssymb}
\usepackage{amsmath}
\usepackage{txfonts}
\usepackage{mathdots}
\usepackage[classicReIm]{kpfonts}
\usepackage[dvips]{graphicx} 
\usepackage[a4paper, portrait, margin=1in]{geometry}
\usepackage{tabularx}
\usepackage{longtable}
\begin{document}


\textbf{\Large Knowledge-based Radiation Treatment Planning: A Data-driven Method Survey}

\textbf{ }

Shadab Momin${}$, Yabo Fu${}$, Yang Lei${}$, Justin Roper${}$, Jeffrey D. Bradley${}$, Walter J. Curran${}$, Tian Liu${}$ and Xiaofeng Yang${}$*

${}$Department of Radiation Oncology, Emory University, Atlanta, GA

\noindent 
\bigbreak
\bigbreak
\bigbreak

\textbf{Corresponding author: }

Xiaofeng Yang, PhD

Department of Radiation Oncology

Emory University School of Medicine

1365 Clifton Road NE

Atlanta, GA 30322

E-mail: xiaofeng.yang@emory.edu

\bigbreak
\bigbreak
\bigbreak
\bigbreak
\bigbreak
\bigbreak

\textbf{Abstract}

This paper surveys the data-driven dose prediction approaches introduced for knowledge-based planning (KBP) in the last decade. These methods were classified into two major categories according to their methods and techniques of utilizing previous knowledge: traditional KBP methods and deep-learning-based methods. Previous studies that required geometric or anatomical features to either find the best matched case(s) from repository of previously delivered treatment plans or build prediction models were included in traditional methods category, whereas deep-learning-based methods included studies that trained neural networks to make dose prediction. A comprehensive review of each category is presented, highlighting key parameters, methods, and their outlooks in terms of dose prediction over the years. We separated the cited works according to the framework and cancer site in each category. Finally, we briefly discuss the performance of both traditional KBP methods and deep-learning-based methods, and future trends of both data-driven KBP approaches.

\noindent \eject 

\noindent 
\section{ Introduction}

Cancer is the second-leading cause of death in North America with the most common types being the cancer of lung, breast, and prostate \cite{RN86}. Radiation therapy (RT), chemotherapy, surgery or their combination are used to control the disease. An approximately 50\% of all cancer patients undergo RT during the course of their illness \cite{RN87}, which makes RT a crucial component of all cancer treatments. In terms of clinical usefulness and effectiveness of RT treatments, the transition from conformal RT to intensity modulated radiation therapy (IMRT) has significantly improved the two-fold dosimetric goal of improving the organ-at-risk (OAR) sparing while maintaining target dose homogeneity and conformity. Furthermore, algorithmic advancements have also played major roles in enhancing the efficiency of RT treatments. These include transition from forward treatment planning to inverse treatment planning approaches, and extension of IMRT to volumetric modulated arc therapy (VMAT). However, despite use of complex inverse optimization algorithms, an inverse planning approach typically demands a large amount of manual intervention to generate a high-quality treatment plan with a desired dose distribution, taking up to a few days before patient gets the first fraction of RT treatment.  
To further enhance the treatment planning efficiency, there has been significant progress into the development of data-driven treatment planning approaches that entail utilizing the knowledge from the past to predict the outcome of a similar, yet new, task. In treatment planning, this concept was introduced by the researchers over a decade ago in the form of knowledge-based planning (KBP). It entails utilizing a large number of previously optimized plans to build a mathematical model or atlas-based repository that can be used to predict the dosimetry (i.e., dose-volume metrics, dose volume histogram (DVH), spatial dose distribution, etc.) for a new patient plan. In 2014, one of the traditional KBP approaches was also made commercially available as RapidPlanTM by the Varian Eclipse® treatment planning system (Varian Medical Systems, Palo Alto, CA). In the past few years, another data driven approach - namely deep learning (DL) has been gaining popularity in the field of radiation oncology for outperforming many state-of-art techniques \cite{RN157, RN156, RN158, RN159, RN163, RN160, RN164, RN161, RN162, RN165, RN166, RN167, RN168, RN169, RN171, RN170}. For instance, convolutional neural network (CNN), a class of deep neural networks (DNN) with regularized multilayer perceptron, have significantly enhanced the performance of imaging and vision tasks. A complex architecture originally designed for image segmentation, also known as U-Net \cite{RN92}, has recently been shown to predict dose distribution without going through a treatment planning process \cite{RN3, RN5, RN111, RN261}.
Though there is a review paper summarizing the articles on traditional KBP methods published between 2011 and 2018  \cite{RN281}, to our knowledge, there is no review paper specific to data-driven dose prediction approaches including both, traditional and recently introduced DL-based KBP. A key difference between traditional and DL-based KBP is the way in which previous knowledge is utilized. In general, traditional KBP methods require user to utilize geometric features such as overlapping volume information between planning target volume (PTV) and neighboring OARs in order to either find the best matched case(s) from repository of previously delivered treatment plans or build dose prediction models (i.e. machine learning (ML), statistical model) \cite{RN56}. DL methods, on the other hand, can learn patterns hidden within the raw data without any requirement of manual feature extraction process, which makes it a more appealing KBP technique compared to the traditional KBP methods. It is important note here that ML-based approaches are included in traditional KBP category in this review as it follows the similar framework to other traditional KBP methods in terms of input (geometric features) and outputs (dose volume metric or DVH). 
The traditional KBP methods include atlas based, statistical modelling and ML methods. While a previous review summarizes these traditional approaches based on methods, current work presents the review of recently emerging DL-based methods as well as the traditional KBP methods from the standpoint of various key parameters and their influence on dose prediction tasks. The goal of this review paper, therefore, is to present the success of traditional KBP methods thus far and highlight the potential of recent DL-based methods in dose prediction tasks. We separate data-driven treatment planning approaches in this regard into two categories: traditional KBP methods and DL-based methods. For each category, we first present a review of key parameters and methods. Subsequently, we present a review of specific investigations and influence of various parameters on dose prediction. Finally, we discuss the advantages and challenges of each dose prediction technique followed by highlighting the potential future trends in data-driven dose prediction methods.

\noindent 

\noindent 
\section{Methods}

\noindent 
\subsection{Literature search}

We searched papers using Elsevier Scopus®, Web of Science, PubMed, Google Scholar and medical physics category of arXiv.org by using logical statements that included the following keywords: knowledge-based treatment planning, ML, DL, dose prediction, RapidPlan®, treatment planning automation, artificial neural network (ANN), convolutional neural network and generative adversarial network. 

\noindent 
\subsection{Article selection criteria}

Only peer reviewed research articles were included in this review. Each research article during literature search was manually scanned based on the information presented in the abstract, which was followed by further in-depth review of specific articles. The articles with description of methodology and comparable or improved aspects of dose prediction quality or efficiency were included. Retrospective studies based on a commercialized KBP approach, RapidPlan®, were also considered. The articles on external beam radiation therapy (IMRT, VMAT, Tomotherapy, Proton etc.…) were included, whereas articles on brachytherapy were excluded. The review of articles on predictions of patient specific quality assurances of a treatment plan was not presented. In this review paper, the term dose prediction includes prediction of entire DVH curve, dose metric (i.e. dose-volume parameter, mean or max dose), voxel dose, spatial dose distribution including slice by slice in 2D manner or 3D dose distribution,  objective weights/constraints based on previous knowledge and also the transfer of all these metrics to the new case for generating an actual plan. 
Figure 1 shows the number of publications per year as well as cumulative publications for both traditional KBP and DL-based dose predictions. Between 2009 and 2014, there was a gradual increase in the number publications on traditional KBP dose prediction in what appears to be the initial development stage of the data driven treatment planning. The curve demonstrates an uplift in the number of traditional KBP articles between 2015 and 2018. Majority of traditional KBP studies in the past few years have been based on commercial RapidPlanTM versus on further expansion of earlier ML or statistical methods. This is certainly not because traditional methods have been fully explored that it has reached its capacity in exploring potential research, but presumably due to recent emergence of DL-based methods owing to their flexibility and superior performances compared to many state-of-the-art techniques. 
In the past few years, the number of publications on DL-based image processing has increased exponentially. To expand the horizons of DL-based applications, researchers have already begun to explore its potential scope for dose prediction tasks. In last four years, the number of DL-based dose prediction publications has gone from 1 in 2016 to already 15 in 2020 as can be seen in Figure 1. The trend appears to demonstrate an increased rate of publications on DL-based versus traditional KBP in the current year.

\begin{figure}
\centering
\noindent \includegraphics*[width=6.50in, height=4.20in, keepaspectratio=true]{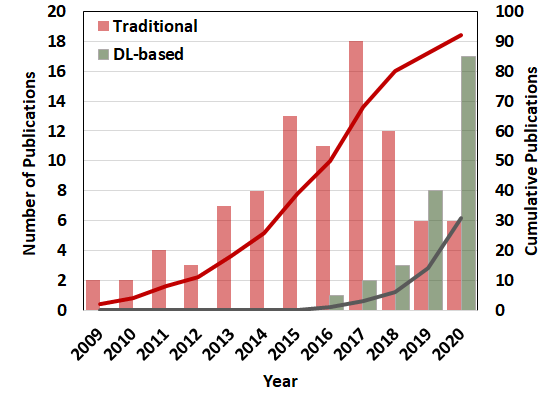}

\noindent Figure 1. The number of dose prediction publications on traditional and DL-based KBP methods per year (bar) with cumulative number of publications (lines).
\end{figure}

\noindent 
\noindent \textbf{Table }\textit{1}\textbf{. }Traditional KBP studies that aimed to predict dose volume histograms (DVHs) for providing a starting point for the plan optimization process.\textbf{ }

\begin{longtable}{|p{0.5in}|p{0.5in}|p{0.7in}|p{0.8in}|p{2.2in}|} \hline 
\textbf{Ref.} & \textbf{Method} & \textbf{Approach/\newline Model} & \textbf{Key Parameters } & \textbf{ Purpose} \\ \hline 
\cite{RN80} & MB & Support Vector Regression & Organ volumes, shape and DTH & To model functional relationship between DVH and patient anatomical shape information. \\ \hline 
\cite{RN18} & MB & Fitting using least square min. & OAR distance to PTV & To translate key parameter correlation to mathematical relationships between OAR geometry and expected dose.   \\ \hline 
\cite{RN37} & MB & Stepwise multiple regression & DTH & To build feature models to identify variation of anatomical features contributing to OAR dose sparing. \\ \hline 
\cite{RN53} & MB & Stepwise multiple regression  & Target, OARs, overlap volumes and DTH & Extension of Yuan \textit{et al.} for intra-treatment-modality model (IMRT -- Tomotherapy)  \\ \hline 
\cite{RN57} & MB & Stepwise multiple regression \newline  & Target, OARs, overlap volumes and DTH, fraction of OAR outside treatment field  & To build two predictive models (single-sparing and standard model) to characterize the dependence of parotid dose sparing on patient anatomical features in summed (primary + boost) plan, rather than two completely separate models.  \\ \hline 
\cite{RN56} & AB & Based on iterative ML algorithm  & Overlapping volume  & To select a reference plan from a library of clinically approved/delivered plans with similar medical conditions and geometry \newline  \\ \hline 
\cite{RN61} & AB & Direct & PTV shape, volume, three spherical coordinates of PTV with respect to OAR OVH & To develop knowledge driven decision support system to assist clinicians to pick plan parameters and assess radiation dose distribution for a perspective patient \\ \hline 
\cite{RN332} & MB & Kernel Density Estimate  & Distance to PTV  & To develop an automated treatment planning solution that iteratively\newline  optimize training set \newline  predicts DVHs for OARs\newline  generates clinically acceptable plans \\ \hline 
\cite{RN237} & MB & Ensemble & Anatomical features, DTH & To combine strengths of various linear regression models to build a more robust model  \\ \hline 
\cite{RN79} & MB & K-nearest neighbors & Generalized-DTH & To characterize DVH variance in multiple target plans \\ \hline 
\multicolumn{4}{|p{1in}|}{\cite{RN63, RN60, RN77, RN256, RN31, RN329, RN149, RN284, RN327, RN94, RN283, RN55, RN236, RN286, RN83, RN62, RN85, RN146, RN64, RN276, RN277, RN59, RN278, RN138, RN143, RN257, RN288, RN97, RN235, RN104, RN21, RN67} \cite{RN331}} & RapidPlan${}^{TM\ }$Eclipse${}^{\circledR }$ treatment planning software:\newline Algorithm is divided in two components: 1) Model configuration and 2) DVH estimation\newline  Mode configuration is divided into data extraction phase and model training phase\newline  DVH estimation consists of DVH estimation phase and objective generation phase  \\ \hline 
\end{longtable}

OVH = overlap volume histogram; DTH = distance-to-target histogram; AB = atlas based; MB = model based 

\noindent 
\section{Results}

\noindent 
\subsection{Knowledge Based Planning}

This review includes over 90 articles on traditional dose prediction methods. These traditional KBP approaches can be classified into two categories: I) Atlas based II) Model based. 
In atlas-based approaches, a physical parameter (i.e. overlap volume histogram (OVH), beam’s eye view projections, tumor location, etc.) is first identified to determine similarity between previous patients’ plans and a new patient plan. This is followed by transfer of knowledge (i.e. dose constraints, DVH values, beam geometrical parameters, DVHs of best matched cases) to predict achievable DVHs or to provide a better starting point to a treatment planner for further trial-and-error optimization. Within atlas-based methods, an indirect approach first predicts the dosimetric parameters through models and features, which are then used to select matching cases. Whereas a direct approach directly predicts a similarity parameter based on features of the plan, CT images, beam’s eye view (BEV) projections. 
In model-based approaches, statistical or ML models are built from previously approved treatment plans. These methods require manually handcrafted features such as PTV-OAR overlap volume, OVH values, OAR distance-to-PTV to predict DVH by using different regression models. 
In this review, we categorized traditional KBP dose prediction articles into three groups according to prediction of: I) entire DVHs in Table 1, II) one or more dose volume metrics in Table 2, and III) voxel doses in Table 3. The articles listed in Table 1 aim to predict the entire DVH for new patient case and utilize the predicted DVHs to guide the treatment planning for a new patient. Commercially available RapidPlanTM module also estimates DVH metrics and generates objectives for a new plan, which are also included in Table 1. Table 2 shows the articles that aim to predict one or more dose metric in order to guide the treatment planning for a new case. Table 3 shows the publications that aim to predict the voxel-level dose distributions to either assist in optimizing a new plan or automatically generate an actual new plan. Figure 2 demonstrates the total number of investigations on traditional KBP methods for various treatment sites. Prostate, head/neck and lung cancers were amongst the most frequently investigated cancer sites as anticipated, whereas very few investigations are conducted on complex sites such as abdominal, intracranial and thoracic. 
In this section, we first provide an overview of key concepts involved in traditional KBP methods. Subsequently, we present a review of different metrics and their extension over the years. Finally, we summarize the influence of different parameters on the performance of traditional methods in dose prediction tasks. 

3.1.1 Dimensionality reduction \\
Though it is desirable to have more data for implementing different models, some implications of having too many data is that they can be redundant, irrelevant, and may result in overfitting, reducing model’s generalizability. Therefore, dimensionality reduction methods were used in majority of traditional KBP studies to decrease the number of variables. Two main components in the process of dimensionality reduction are: feature extraction and feature selection. The process of feature extraction begins with an initial set of features followed by redefining with the intention for them to be more informative. Principle component analysis (PCA) is one of the most commonly used reduced order modeling techniques in model-based approaches. The PCA determines features that retain the most of the variation among the data \cite{RN282} so that they can be represented by a smaller number of dimensions. For example, in a binary classification problem, if the goal is to classify an object A, represented by a P number of features in a P-dimensional vector, as either of two classes. If P is too large, some characteristics may be more valuable than others for the purpose of classification. The goal of PCA is to reduce the dimensionality of the dataset consisting interrelated variables into a smaller set of mutually uncorrelated variables \cite{RN282}. Feature selection process involves the selection of valuable features from the ones at our disposal. In many traditional KBP studies, the PCA is used in the process of feature selections \cite{RN63, RN60, RN77, RN31, RN66, RN55, RN233, RN236, RN62, RN36, RN53, RN277, RN59, RN278, RN143, RN46, RN65, RN235, RN21, RN37, RN57, RN237, RN80}. 

3.1.2 Various features/metrics \\
A common theme in majority of traditional approaches is that the optimality of desired plan is strongly influenced by geometries of critical structures with respect to the PTV. Commonly reported geometric features include OVH, distance to target histograms (DTH), OAR distance-to-PTV. The influence of parotid size and proximity to the PTV on the dosimetric sparing of parotid was first studied by Hunt \textit{et al.} \cite{RN324}. In addition to geometric features, additional plan features such as PTV-OAR volumes, mutual information including beams’ eye view projections, number and angles of specified beams and photon energy have also been utilized in traditional KBP studies. List of these key parameters are tabulated in Table 1, 2 and 3 along with their corresponding references.  

\textbf{Overlap volume histogram (OVH) based methods}

The OVH was introduced to study the influence of OAR’s proximity to the target on its received dose. It is one of the most frequently used metrics in both atlas-based and model-based approaches as can be seen in Table 1 and 2. Wu \textit{et al.} and Kazhdan \textit{et al.} first shed light on the concept of the OVH as a one-dimensional function measuring the proximity of an OAR to the target \cite{RN101, RN43}. The OVH calculation involves uniform expansion and contraction of the target. Target contraction and expansion is repeated until OAR completely overlaps the target and there is no overlap between the target and the OAR, respectively \cite{RN43}. In other words, it is the percentage of the OAR’s volume that overlaps with a uniformly expanded or contracted target. In general, OVH-driven models assume that the dose to an OAR is inversely proportional to its distance from the target. 

A large array of studies has combined historical data with the OVH methods for prediction of entire DVH (Table 1) and one or more dose metrics (Table 2). Wu \textit{et al.} used the OVH for its use in head/neck IMRT treatment plan quality control to help planners with evaluation \cite{RN43}. This was followed by using OVH to generate the achievable DVH objectives for head and neck cancer case \cite{RN52}. With a model based on OVH \cite{RN101} and PCA \cite{RN37, RN80}, Wang \textit{et al.} investigated the effect of interorgan dependency and impact of data inconsistency \cite{RN97}. Larger prediction errors were found for head/neck region (<4 Gy for 83\% of the cases) compared to similar model applied to prostate case (<2 Gy for 96\%) presumably due to interorgan dependency \cite{RN97}. Moore \textit{et al.} also used OVH information to predict OAR dose metrics for head/neck and prostate IMRT plans \cite{RN98}.  Yuan \textit{et al.} used OVH metric to quantify the effects of an array of patient anatomical features of the PTV and OARs and their spatial relationship on interpatient OAR dose sparing in IMRT and found mean distance between OAR and PTV, mean volume between OAR and PTV, out-of-field volume of OARs and geometric relationship between multiple OARs to be important factors contributing to the organ dose sparing \cite{RN37}. For multiple OARs, using separate OAR-specific prediction models was found to be more accurate in predicting voxel doses compared to all OAR voxels in a single training model \cite{RN75}.
 
\begin{figure}
\centering
\noindent \includegraphics*[width=6.50in, height=4.20in, keepaspectratio=true]{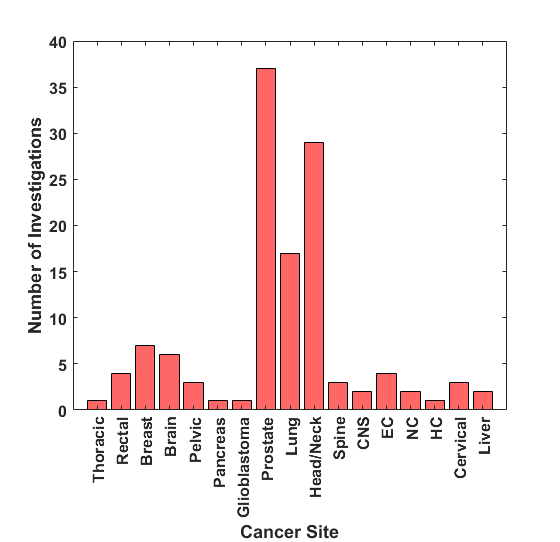}

\noindent EC = Esophageal cancer; NC = Nasopharyngeal carcinoma; HC = Hepatocellular Cancer\\
 Figure 2. The total number of traditional KBP investigations on dose prediction for various cancer sites.
\end{figure}

The success of OVH based prediction primarily rests on the observation that the minimum achievable dose to OAR depends on its distance and orientation to the PTV. However, the OVH based model \cite{RN52} has been investigated for pancreatic cancer in which the OARs are larger compared to the tumor, part of OARs can engulf the PTV, and highly deformable organs can vary the beam configurations among different patients \cite{RN173}. Petit \textit{et al.} showed that the OVH based predicted doses were achieved within 1 and 2 Gy for more than 82\% and 94\% of the patients, respectively, with improvement of 1.4 Gy and 1.7 Gy for mean dose to the liver and kidneys, respectively. To further investigate the capability of OVH parameter, the global shift of the OVH was quantified after hydrogel injection to represent the efficacy of hydrogel injection in separating the rectum from PTV. The OVH was found to be a better metric for rectum sparing than the hydrogel volume \cite{RN19}. Wang \textit{et al.} used OVH to build a treatment planning QA model from consistently planned pareto-optimal plans for prostate cancer, improving planning standardization and preventing validation with possibly suboptimal benchmark plans \cite{RN274}. 
In earlier OVH-driven studies, a large variations in IMRT dose at a given OVH distance for a specific fractional volume of an OAR was reported \cite{RN279, RN37}. To address this disparity in the distance-to-dose correlation, Wall \textit{et al.} studied inherent inter-planner variations in plan quality of the previous plans and second order dosimetric and anatomical factors. Out of all factors, in-field bladder and rectal volume showed the strongest correlation (R = 0.86 and R = 0.76) with doses. Therefore, in-field OAR volume was incorporated into the OVH only metric \cite{RN51}.  Generic OVH introduced by Kazhdan \textit{et al.} directly infers a DVH rather than a spatial dose distribution \cite{RN101}. With multi-patient atlas based-dose prediction approach, McIntosh and Purdie demonstrated that incorporating spatial information into the model can improve the dose prediction accuracy in comparison to the generic OVH method. This method was found to be less important for breast cavity and lung whereas improved prediction accuracy for whole breast, rectum and prostate cancer \cite{RN244}. 

\noindent \textbf{Table 2. }Traditional KBP studies that aimed to predict one or more dose metrics for providing a starting point for the plan optimization process.\textbf{}

\begin{longtable}{|p{0.6in}|p{0.5in}|p{0.6in}|p{0.8in}|p{2.2in}|} \hline 
\textbf{Ref.} & \textbf{Method} & \textbf{Approach/\newline Model} & \textbf{Key Parameters} & \textbf{Purpose} \\ \hline 
\cite{RN150} & MB & Support vector regression & OAR, DV constraint settings & To create an accurate IMRT plan surface as a decision support tool to aid treatment planners  \\ \hline 
\cite{RN240} & AB & Direct & Clinical stage, and gleason score & To update the weights of difference clinical parameters for a new patient through a group based simulated annealing approach  \\ \hline 
\cite{RN52} & AB & Direct & OVH & To use geometric and dosimetric information retrieved from a database of previous plans to predict clinically achievable dose volume metric \textit{(A retrospective based on method by }{\textbackslash}cite$\{$RN43$\}$\textit{)} \\ \hline 
\cite{RN241} & AB & Direct & OVH & To implement OVH-based automated planning system to improve quality, efficiency and consistency for head and neck cancer \\ \hline 
\cite{RN99} & AB & Direct & OVH & To predict dose to 35\% of rectal volume as a treatment planning quality assurance for prostate cancer patients. \\ \hline 
\cite{RN144} & AB & Direct & OVH & To investigate if OVH driven IMRT database can guide and automate VMAT planning for head and neck cancer \\ \hline 
\cite{RN19} & MB & Linear Regression & OVH & To evaluate OVH metric for prediction of rectal dose following hydrogel injection  \\ \hline 
\cite{RN46} & MB & Stepwise multiple regression & OVH & To utilize patients anatomic and dosimetric features to predict the pareto front\newline  \\ \hline 
\cite{RN242} & MB & Logistic Regression & Distance to the tangent field edge & To predict left anterior descending artery maximum dose. Model to guide the positioning of the tangent field to keep maximum dose $\mathrm{<}$ 10 Gy \\ \hline 
\cite{RN333} & MB & Linear Regression & OAR volumes  & To develop a model to predict attainable prescription dose for IMRT of entire hemithoracic pleura  \\ \hline 
\cite{RN45} & MB & Curve Fitting & Rectum-target overlap & To predict optimum average rectum dose \\ \hline 
\cite{RN50} & MB & Stepwise Regression & Target OAR overlap & To predict mean parotid dose \\ \hline 
\cite{RN51} & AB & Direct & OVH', In field OAR volumes & The minimum DVH value at the percentage volume of the bladder and rectum was used \\ \hline 
\end{longtable}

OVH = overlap volume histogram; DTH = distance-to-target histogram; AB = atlas based; MB = model based

\textbf{Projection based methods}

These algorithms typically rely on matching 2D images, beam’s eye view (BEV) of the projection at each corresponding gantry angle, based on statistical properties of image histogram. The best matched case is generally identified based on the sum of mutual information (i.e., beams eye view projection) values for each of the total number of beam angles involved. This method has been used for prostate \cite{RN76} and head/neck cancer \cite{RN153}. Good \textit{et al.} calculated mutual information representing the best match for the query case. The PTV projection of matched case were deformed to the query case’s PTV projections at each angle to adjust for shape differences between the PTVs of the query and match case. This approach reduced doses to the OARs and improve target dose conformity and homogeneity in KBP generated plans compared to the original plans \cite{RN68}. 

\textbf{Distance-to-target histogram (DTH) based methods}

Distance to target histogram (DTH) is the fractional volume of the OAR within certain distance from the PTV surface. This metric along with volumes of the PTV and OARs are typically used as input features in ML approaches such as multivariable nonlinear regression (MVNLR) and support vector regression (SVR) \cite{RN80}. It is important to note that DTH is equivalent to OVH \cite{RN101} when the Euclidean form of the distance function is used. This DTH metric was extended to generalized distance-to-target histogram (gDTH) by Zheng \textit{et al.} in order to account for the relative shape distribution of multiple PTVs for head and neck cancer \cite{RN79}. In comparison to conventional model, the gDTH model improved DVH prediction accuracy for brainstem, cord, larynx, mandible, parotid, oral cavity and pharynx \cite{RN79}. While this gDTH model presented similar plans with respect to an individual OAR, to develop a knowledge based tradeoff hyperplane model that assists with clinical decisions, the concept of gDTH was further extended to select similar plans with respect to all OARs by employing case similarity metric that is a weighted sum of gDTH Euclidean distances between two cases across all OARs \cite{RN258}. Finally, the DTH has also been utilized with multivariate regression-based models, which is commercially available as RapidPlanTM in Eclipse® treatment planning software.

3.1.3. Influence of various parameters \\

\textbf{Outliers/Data inconsistency}

Outlier detection is one of the important factors to consider when building a data driven dose prediction model or repository that is generalizable to new cases. Outliers can reduce the goodness of fit between geometry and dosimetry, which, in turn, can comprise the model performance \cite{RN98}. Two commonly reported outliers in the literature are geometric outliers and dosimetric outliers. Geometric outliers, on the other hand, entail large anatomical variations including OAR distance to the PTV. An example of geometric outlier is including a prostate + nodes case to prostate only cases. Several studies investigated the influence of outliers on model performance as shown in Table 4. Dosimetric outliers represent the presence of plans in which OARs are not actively spared or there are violations of dose-volume criteria. In other words, dosimetric outliers are the plans for which the re-planning can significantly reduce in OAR dose without compromising target coverage. Appenzoller \textit{et al.} described a model to identify outliers in the form of suboptimal plans and showed that excluding outliers in refined model resulted in a strong correlation between predicted and realized gains after re-planning (r = 0.92 for rectum, r = 0.88 for bladder and r = 0.84 for parotid glands). For head/neck RapidPlanTM based KBP, Delaney \textit{et al.} analyzed the influence dosimetric outliers and showed a moderate degradation in accuracy of the model attributed to decreased precision of the estimated DVHs \cite{RN284}. 
For pelvic cases, Sheng \textit{et al.} assessed the effectiveness of outlier identification by studying the impact of both, geometric and dosimetric, outliers. This study suggested a greater impact of dosimetric outliers with negative impact on both bladder and rectum model compared to geometric outliers with negative impact only on bladder model \cite{RN136}. Wang \textit{et al.} studied effect of data inconsistency with respect to planning prioritizations through a) mixed training dataset with a consistent validation dataset b) a consistent training dataset with a mixed validation dataset c) both a mixed training and validation dataset d) both consistent training and validation dataset and found that data inconsistency led to a large increase in prediction error with errord < errorc < errora < errorb. \cite{RN97}. In addition to removing the outliers (i.e. suboptimal plans) from the training cohort \cite{RN18}, an alternative to address the issue of outliers reported in the literature is re-planning of the identified suboptimal plans for prostate and head/neck cancer \cite{RN325} and lung cancer \cite{RN85} followed by inclusion in the training cohort. Clinically available RapidPlanTM provides different statistical evaluation metrics for identifying the outliers as shown in Table 4.

\textbf{Diversities within traditional methods}

Many retrospective studies were published after 2014 presumably due to clinical implementation of traditional KBP module in the form of RapidPlanTM in Eclipse® treatment planning software. These studies investigated the applicability of traditional methods with respect to variations in external parameters (i.e. multi-modality, multi-institution, sample size etc.). Here, we present a review of these studies with their findings. Wu et al 2013 used the DVH objectives derived from previous IMRT plans as an optimization parameter for VMAT treatment planning in head/neck cancer, resulting in a similar dosimetric quality compared to IMRT plans \cite{RN144}. Wu et al demonstrated that supine VMAT model for rectal plans can optimize IMRT plans of prone patients, yielding superior OAR sparing and quality consistency than conventional treatment planning method \cite{RN104}. The prediction models trained on Helical Tomotherapy for prostate cancer were utilized to predict constraints to perform an optimization of new plans using RapidArcTM technique, it resulted in comparable/increased bladder and rectum doses compared to expert planner’s plan. Delaney \textit{et al.} demonstrated that using a model only on photon beam characteristics could make the DVH predictions for proton therapy and can be used as a patient selection tool for protons \cite{RN149}. McIntosh \textit{et al.} studied contextual atlas random forest (cARF) algorithm with and without OAR region of interest features and found that the algorithm can pick better atlases without ROI features, however is not compatible enough to map the dose distribution from those atlases onto a new patient \cite{RN69}.
Huang et al demonstrated that RapidPlanTM model for one energy (10 MV) can generate dose volume objectives for plans with 6 and 10 MV, however a  RapidPlanTM  model for flattened beams cannot optimize un-flattened beams prior to adjusting the target objectives \cite{RN83}. A RapidPlanTM module also has the potential to generate high quality treatment plans on a newly implemented treatment planning software compared to manually optimized plans for prostate cancer \cite{RN276}. For esophageal cancers, the RapidPlan created from plans optimized using RayStationTM produced comparable lung doses \cite{RN257}. 
For patients enrolled in Radiation Therapy Oncology Group (RTOG) 0617, Kavanaugh et al showed the feasibility of a single-institution RapidPlanTM model as a quality control tool for multi institutional clinical trials to improve overall plan quality and provide decision support to determine the need for clinical trade-offs between target coverage and OAR sparing \cite{RN85}. For prostate cancer, Schubert \textit{et al.} have demonstrated the possibility of sharing models among different institutes in a cooperative framework \cite{RN277}. For prostate cancer RapidPlanTM amongst five different institutions, Ueda \textit{et al.} also suggested that it is critical to ensure similarity of the registered DVH curves in the models to the institution’s plan design before sharing the models. For prostate cancer, Good \textit{et al.}, applied the model trained on their institute to generate plans for patient datasets outside institution with the potential for homogenizing plan quality by transferring planning expertise from more to less experienced institutions \cite{RN68}. Good \textit{et al.} achieved superior or equivalent to the original plan in 95\% of 55 tests patients \cite{RN68}. More recently, a disease site specific multi-institutional, NRG-HN001 clinical trial based RapidPlanTM model was built as an offline quality assurance tool for which it improved sparing of OARs in a large number of reoptimized plans submitted to the NRG-HN001 clinical trial \cite{RN286}.

\textbf{Sample size}

Figure 3 shows an average number of training and test set for each cancer site in traditional KBP methods with standard deviation over number of investigations listed on the top x-axis. The number of training/test sample size were not directly mentioned or required in the methods described in some publications.
For RapidPlanTM , it is indicated that the minimum number of plans required for model creation is 20, however adding additional plans will usually help create a more robust plan \cite{RN328}. Numerous studies have compared the quality of plans generated by RapidPlan by using high quality plans in training and found that 25 – 30 plans may produce clinically acceptable plan for prostate \cite{RN283} and head/neck \cite{RN278} cancer. For prostate cancer, Boutilier et al analyzed effects of the training set size on the accuracy of four models from three different classes: DVH point prediction, DVH curve prediction and objective function weights. The authors concluded that minimum required sample size depends on the specific model and endpoint to be predicted \cite{RN95}.  Zhang et al showed that approximately 30 plans were sufficient to predict dose-volume levels with less than 3\% relative error in both head and neck and whole pelvis/prostate \cite{RN150}.

The requirement of sample size also partially depends on the robustness of the model used. Yuan \textit{et al.} used 64 and 82 cases for prostate and head/neck case, respectively, in support vector regression (SVR) model for DVH predictions \cite{RN37}. Landers \textit{et al.} demonstrated statistical voxel dose learning (SVDL) to be more robust to patient variability compared to spectral regression and SVR for noncoplanar IMRT and VMAT for head/neck, lung and prostate cancer by using 20 cases for each site in 4-fold cross-validation \cite{RN75}. An atlas-based dose prediction \cite{RN244} is more sophisticated method in which each patient in the training set represents 1 atlas. Feature extraction and characterization is typically performed on CT of the patients, which results in a probabilistic dose estimates to find the most likely voxel dose from similar atlases. In comparison to ANN and SVR methods, a large training sample sizes were required for this method (58 for rectal, 77 for lung, 97 for breast cavity, 113 for central nervous system (CNS) brain, 144 for breast and 144 for prostate cancer).
Overall, the review of traditional KBP dose prediction publications thus far suggests an improved efficiency compared to manual optimization, sufficient flexibility of traditional KBP methods in terms of their applicability (i.e. multimodality in EBRT), the need of these models for more complex sites, the requirement of an automated approach for accounting for outliers to further enhance the treatment planning efficiency and the potential of building site specific universal RapidPlanTM models for multi-institution adaptation. 

\begin{figure}
\centering
\noindent \includegraphics*[width=6.50in, height=4.20in, keepaspectratio=true]{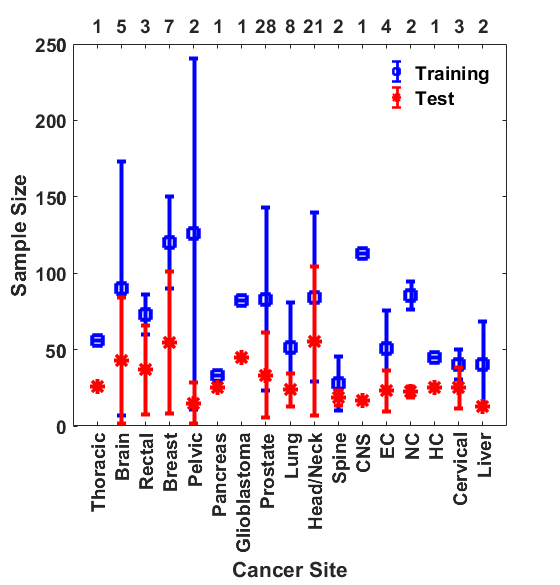}

\noindent CNS = Central Nervous System; NC = Nasopharyngeal Cancer; EC = Esophageal Cancer \\
Figure 3. The average number of training and testing datasets in traditional KBP dose prediction methods for each cancer site. The values are averaged over number of investigations listed on top x-axis and the error bars represent standard deviation.
\end{figure}

\noindent 
\subsection{Deep Learning}

DL offers numerous advantages and support to personals of different disciplines in the different steps of radiotherapy treatment planning. An appealing feature of DL methods is that the layers of features are not manually designed, rather learned directly from raw data. Because DL methods are good at discovering intricate structures in high-dimensional data, it is applicable to a wide range of applications in science  \cite{RN232}. In this section, we provide an overview of different architectures and neural networks that have been applied to dose prediction tasks up to now. 
The use of DL in dose prediction was initially utilized in the form of  ANN \cite{RN34}. In these earlier DL-based methods, organ volumes including PTV and OARs, number of fields and distances from OARs to the PTV were used to train ANN, which was then used to correlate dose at a given voxel to a number of geometric and plan parameters, similar to that of used in traditional KBP methods. The DNNs are the most commonly used networks in DL-based dose prediction. It resembles the traditional ANN, but with a large number of layers. Therefore, ANN-based studies are included into DL-based dose prediction category in Table 5 despite their comparable framework to that for traditional KBP methods.  Neurons within each layer are nodes which are connected to subsequent nodes via links that correspond to biological axon-synapse-dendrite connections, analogous to the neural cell of human. The layers embedded between an initial input layer and the final output layer are called hidden layers. The number of layers determines network’s width, whereas the number of neurons determines its depth. Each neuron between its input and output undergoes a linear followed by a non-linear operation. 
In layered format, each neuron receives the information from the neurons in the previous layer and passes it to neurons of the next layer after processing it. On the other hand, the residual connections can be added to connect neurons in non-adjacent layers such as ResNet proposed by He \textit{et al.} \cite{RN117}. The ResNet architecture has been presented with different number of layers: ResNet (18, 34, 50, 101, 152). Many DNN architectures have been presented for various applications. For dose predictions, CNN – namely fully convolutional neural network (FCN) and fully connected CNN (FCNN) have been used so far. A DL-based generative model, commonly known as generative adversarial network (GAN), has also been employed to aid the main network (FCN) for predicting dose distribution. 

3.2.1 Convolutional Neural Network \\
Multilayer perceptron has the fully connected networks in which each neuron in one layer is connected to all the neurons in the next layer. It is now succeeded by CNN, a class of DNN with regularized multilayer perceptron \cite{RN29}. CNN, by far, is the most widely used DNN for dose prediction task as can be seen in Table 5. Main components of a typical CNN are convolutional layers, max pooling layers, batch normalization, dropout layers, a sigmoid or softmax layer. 
The convolutional layer consists of a set of convolutional kernels where each kernel acts as a filter. The image is divided into small slices, known as receptive fields, through convolutional kernel, which aids in extracting features. Kernel uses a specific set of weights to convolve with corresponding elements of the receptive field. The weight sharing ability of convolutional operation allows extraction of different set of features within an image by sliding kernel with the same set of weights on the image. This makes CNN parameter more efficient compared to the fully connected networks. This operation can be grouped based on the type and size of filters, direction of convolution, and type of padding \cite{RN232}. 
From the result of convolution operation, the feature motifs can occur at different locations in the image. The goal is to preserve its approximate position relative to others rather than the exact location. The pooling or down-sampling sums up similar information in the neighborhood of the receptive fields and outputs the dominant response within this local region, helping to extract combination of features that are invariant to translational shifts \cite{RN291}. Commonly reported pooling formulations used in CNN are max, average, L2, spatial pyramid pooling and overlapping \cite{RN293, RN294}
Nonlinear operation, also known as activation function, helps in learning of sophisticated patterns by serving as a decision function. Different activation functions reported in the literature are sigmoid, tanh, SWISH, ReLU and its variants including leaky-ReLU, PReLU have been used to inculcate non-linear combination of features \cite{RN298, RN295, RN297, RN294, RN296}. More recently proposed activation function is MISH, which has shown better performance than ReLU on benchmark datasets \cite{RN299}. 
Batch normalization is applied to address the question of internal covariance shifts, a change in the distribution of hidden unit values, within feature maps that can reduce the convergence speed. It essentially unifies the distribution of feature map values by setting them to zero mean and unit variance, which, in turn, improves the generalization of the network by smoothening the flow of the gradient \cite{RN300}. 
Finally, weight regularization and dropout layers are used to alleviate data overfitting. The difference between the predicted and the target output is calculated through loss function. CNN is generally trained by minimizing the loss via gradient back propagation using optimization methods. 
Different architectures have been proposed in the literature to enhance the performance of CNN. U-Net, originally introduced for segmentation of neuronal structures in electron microscope stacks \cite{RN92}, is the one of the most widely used architectures in CNN. In addition to segmentation, it is also used for image-to-image translation tasks that outputs an image that has a one-to-one voxel correspondence with the input. U-Net permits effective feature learning even with small number of training sample size. Milletary \textit{et al.} proposed a three dimensional variant of U-Net known as V-Net \cite{RN301}.
A known issue of training DNN occurs from the vanishing gradient. Therefore, ReLU \cite{RN232} and its variants are generally preferred as activation function owing to their ability in overcoming the vanishing gradient problem \cite{RN302}. LeCun \textit{et al.} formulated the layers as learning residual functions instead of directly fitting a desired underlying mapping \cite{RN232}. A densely connected neural network (DenseNet) by Huang \textit{et al.} connects each layer to every other layer \cite{RN303}. More recently, attention gate was used in CNN in order to suppress irrelevant features and highlight salient features useful for a given task \cite{RN304}. 

3.2.2 Generative Adversarial Network \\
Generative adversarial network (GAN) is a widely used semi-supervised learning method in DL \cite{RN305}. Two major components of GAN are generative network and discriminator network that are trained concurrently to compete against each other. The goal of generative network is to generate artificial data that can approximate a target data distribution from a low-dimensional latent space, whereas the goal of discriminator network is to recognize the data presented by the generator and flag it as either real or fake. Both networks get better over the course of training to reach nash equilibrium, which is the minimax loss of the aggregate training protocol \cite{RN305}. Some of popular variants of GAN include CycleGAN \cite{RN307}, conditional GAN (cGAN) \cite{RN306} and StarGAN \cite{RN308}. GAN is widely used in medical imaging \cite{RN159, RN309, RN163, RN160}. 

3.2.3 Reinforcement Learning \\
Reinforcement learning (RL) trains an agent, connected to its environment through perception and action, to make adjustments based on interaction between the agent and the environment. The agent gets certain indication of the current knowledge of the environment at each step of its interaction. Based on received indication, the agent then chooses an action to generate as output. This action changes the state of the environment, the value of this state transition is communicated to the agent through a reward function. The agent’s behavior can learn to do this over time through trial and error \cite{RN342}.  In other words, the goal of RL is to find the balance between the search and the current knowledge. RL has been combined with DNN to accomplish human-level performances \cite{RN343}. RL is a unique framework that resembles the workflow of treatment planning optimization. The potential scope of RL in DL-based dose prediction task (Table 5) has been investigated in a recent study \cite{RN340}. RL was used to train a DNN named virtual treatment planner network, which, in turn, decides the way of changing treatment planning parameters to improve plan quality instead of a treatment planner similar to the treatment planning process \cite{RN340}.

\noindent \textbf{Table 3. }Traditional KBP studies that aimed to predict voxel level doses for providing a starting point for the plan optimization process.

\begin{longtable}{|p{0.5in}|p{0.4in}|p{0.7in}|p{0.8in}|p{2.3in}|} \hline 
\textbf{Ref.} & \textbf{Method} & \textbf{Approach/\newline Model} & \textbf{Key Parameters} & \textbf{Purpose} \\ \hline 
\cite{RN76} & AB & Direct & BEV's projections & To identify similar patient cases by matching 2D BEV projections of contours  \\ \hline 
\cite{RN68} & AB & Direct & BEV's projections  & To adapt matched case's plan parameters from one institute to optimize the query case of an outside institution  \\ \hline 
\cite{RN42, RN41} & MB & Multivariate analysis Slice weight function & Distance-to-PTV,\newline Slice level  & To determine the relationship between the position of voxels and corresponding doses to predict sparing of the OARs  \\ \hline 
\cite{RN141} & MB & Active shape model, active optical flow model  & PTV locations in relation to spinal cord  & To study the effect of PTV contours on dose distribution at spinal cord.\newline  Five subgroups were created according to the PTV locations in relation to spinal cords.  \\ \hline 
\cite{RN153} & AB & Direct & Target-OAR overlap  \newline Shell creation surrounding the match target volume & To adapt the matched case from the database for query case by deforming the match beam fluences, warping the match primary/boost dose distribution and distance scaling factor \\ \hline 
\cite{RN70} & AB & Direct & Target-OAR overlap  & To transfer the beam settings and multileaf collimator positions of best match case to the new case  \\ \hline 
\cite{RN39} & AB & Direct & The PTV and Seminal vesicles (SV) concaveness angle and \% distance from SV to the PTV & To transfer treatment parameters of the atlas case to the new case  \\ \hline 
\cite{RN33} & AB & Indirect & Multi-scale image appearance features  & To use contextual atlas regression forest (cARF) augmented with density estimation over the most informative features to learn an automatic atlas-selection metric for dose prediction  \\ \hline 
\cite{RN244} & AB & Indirect & Features based on the spatial dose distribution and features derived from DVHs & To extend CRF by introducing conditional random field model (cARF-CRF) to transform the probabilistic dose distribution into a scalar dose distribution that adheres to desired DVHs.\newline  \\ \hline 
\cite{RN69} & AB & Indirect & Multi-scale image appearance features & To converts a predicted per voxel dose distribution into a complete radiotherapy plan through fully automated pipeline using cARF-CRF.  \\ \hline 
\end{longtable}

OVH = overlap volume histogram; DTH = distance-to-target histogram; AB = atlas based; MB = model based

\noindent 
\subsection{Deep Learning in Dose Prediction}

DL-based dose prediction methods can be categorized according to DL properties such as network architectures (CNN, GAN etc.), training process (supervised, unsupervised, semi-supervised, deep reinforcement etc.), input image types (CT only, CT + OAR + PTV contours, etc.), output types (2D or 3D dose distribution) and sample size (training, testing etc.). As shown in Figure 1, DL-based dose prediction methods have gained popularity amongst the researchers only in the past few years, there are nearly 30 publications on DL-based dose prediction so far. These DL-based dose prediction publications are tabulated in Table 5 along with their network architectures, input and output characteristics. Figure 4 represents the total number of DL-based dose prediction investigations per treatment site. This follows a similar trend to that observed for traditional KBP dose prediction approaches with the highest number of investigations being on prostate and head/neck cancer sites. Here, we categorized DL-based dose prediction publications thus far into two groups based on network architectures: I) CNN – namely U-Net architecture and II) GAN. We first provide the review of work for each network architecture followed by their applicability on various dose prediction application and limitations. Subsequently, we discuss the influence of different parameters in DL-based dose prediction methods.  

3.3.1 Overview of CNN based works \\
As shown in Table 5, U-Net has been widely used CNN architectures used for predicting dose distributions. U-Net is effective in terms of calculation and combination of global and local features because it is consisted of encoding and decoding path. The decoding path concatenates the features from both previous layers in encoding path and features from current layers in decoding path. Many variants of U-Net including 3D U-Net have appeared in literature for dose prediction purposes. 
Earlier work in DL-based dose prediction methods involved predicting doses in 2D manner \cite{RN3, RN111}. Sumida \textit{et al.} used the U-net model, initially proposed by Ronneberger \textit{et al.} \cite{RN92}, to make 2D dose prediction. Two main flows of this were encoding and decoding parts. Encoding parts layers followed 2D convolution layer, batch normalization, rectified linear unit (ReLU) and max-pooling layer. The network was trained to make dose prediction for Acuros XB (AXB) from low resolution dose calculated through AAA algorithm and CT. Similarly, Nguyen \textit{et al.} also trained a seven-level hierarchy with modified version of original U-Net to make dose prediction for a prostate case \cite{RN111}.

More recent works were focused on predicting 3D dose distributions using DL methods. To overcome increased computation load in 3D dose prediction, Nguyen \textit{et al.} proposed Hierarchically Densely U-Net (HD U-Net), which not only was able to predict 3D dose distribution, but also outperform dose predictions made by standard U-Net model \cite{RN10}.  HD U-Net combines DenseNet’s efficient feature propagation and utilize U-Net’s ability to infer both local and global feature by connecting each layer to every other layer in feed-forward fashion, yielding better RAM usage and better generalization of the model. To further simplify 3D dose prediction problem and increase prediction accuracy, Xing \textit{et al.} projected the fluence maps to the dose distribution using a fast and inexpensive ray-tracing dose calculation algorithm and trained HD U-Net to map the ray-tracing low accuracy dose distribution (does not consider scatter effect) into an accurate dose distribution calculated using collapsed cone convolution/superposition algorithm \cite{RN107}. 
DL-based methods have also been expanded to predict pareto optimal dose distributions so that physicians can learn the desired dosimetric trade-offs in real time and learn the viability of different dosimetric goals. Ma \textit{et al.} constructed 3D U-Net architecture to predict individualized dose distribution for different tradeoffs \cite{RN265}. In predicting pareto dose distribution, the network should be able to map many dose distributions from a single anatomy. In doing so, it should be able to differentiate between the clinical consequences and corresponding predicted dose distribution. To address this clinically relevant differences amongst different dose distribution, Nguyen \textit{et al.} proposed the differentiable loss function based on the DVH and adversarial loss in addition to traditional voxel wise mean square error (MSE) loss to train the network \cite{RN264}. Along the same line of work, Bohara \textit{et al.} incorporated beam information to predict pareto dose distribution using anatomy-beam model proposed by Barragán‐Montero \textit{et al.} \cite{RN108}.

U-Net architecture has also been used for internal radiation dose predictions \cite{RN105, RN267} where the network was trained to predict 3D dose rate maps given the mass density distribution and radioactivity maps. Since clinically available Medical Internal Radiation Dose Committee (MIRD) based dose estimations are least precise, the long-term goal of these studies is to create a stable DL-based dose estimation model that achieves a precision close to that of Monte Carlo simulations. 
He \textit{et al.} proposed residual network, known as ResNet, to mitigate the difficulty of training DNN caused by gradient vanishing \cite{RN117}. He \textit{et al.} reformulated the layers as learning residual function instead of directly fitting a desired underlying mapping. Chen \textit{et al.} and Fan \textit{et al.} proposed DL method based on ResNet with 101 and 50 weight layers, respectively, to predict dose distribution for head/neck cancer IMRT patients \cite{RN9, RN272}. Since networks with very deep layers are difficult to train due to vanishing gradient, such networks used shortcut connections to add to the outputs of the stacked layers \cite{RN117}. More recently, Liu \textit{et al.} proposed ResNet for dose prediction in the nasopharyngeal cancers for Helical Tomotherapy. To achieve multi-scale feature learning, Liu \textit{et al.} divided the ResNet into several parts without fully connected layers and respectively combined with input data to achieve pixel-wise feature abstraction and extraction in structural image. 

3.3.2 Overview of GAN-based works \\
GAN entails a pair of neural networks: a generator and a discriminator. From the treatment planning standpoint, generator could be represented as the treatment planner who generates the plan and radiation oncologist could be represented as discriminative network who evaluates the plan generated by the treatment planner. Both the treatment planner and a radiation oncologist get better at performing their tasks as they become experienced over time. Only a handful of studies have investigated the performance of GAN for dose prediction task as shown in Table 5. 
Mahmood \textit{et al.} demonstrated the first use of 2D GAN for predicting dose for each 2D slice independently for oropharyngeal cancer. Subsequently, Babier \textit{et al.} proposed the first 3D GAN for prediction of full 3D dose distributions, which outperformed the 2D GAN model proposed by Mahmood et al presumably owing to its ability to learn the vertical relationship between adjacent axial slices in contrast to 2D GAN networks.  Recently, Vasant \textit{et al.} proposed a novel 3D attention-gated generative adversarial network (DoseGAN) as a superior alternative to current state of the art dose prediction networks \cite{RN261}. Spatial self-attention allows networks to emphasis portions of the intermediate convolution layers. Attention gated proposed by Vasant \textit{et al.} can potentially offer deeper and more efficient discrimination, while being trained in parallel with the generator network and facilitating the model convergence \cite{RN261}. This addresses the issue of keeping the number of networks parameters as low as possible in conventional GAN. Attention-gated GAN proposed by Vasant \textit{et al.} outperformed conventional 2D and 3D GAN in all dosimetric criteria including PTV and OARs \cite{RN261}. 
All four studies \cite{RN285, RN11, RN6, RN261} on GAN-based dose predictions constructed a generator and discriminator network using the pix2pix architecture proposed by Iosa \textit{et al.} \cite{RN317}.  U-net generator was used, which passes a contoured CT image slice thorough consecutive layers, a bottleneck layer and subsequent deconvolution layers. U-net also uses skip connections to easily pass high dimensional information between the input (CT image slice or contoured structures) and the output (dose slice). 

3.3.3 Overview of learning processes \\
In this section, we briefly present a review of four learning processes including supervised learning (SL), unsupervised-learning (USL) and semi-supervised learning (SSL) that have been utilized so far in DL-based dose prediction tasks. Earlier approaches used SL that trained a model by using labeled data in the form of different geometrical parameters and distance to the target to train the network. In contrast, USL does not require such target information and rely solely on the input data to learn the patterns hidden within raw data. A typical example of USL is training deep auto encoder (DAE), which has a flexible network structure with encoder and decoder. These USL networks can be CNN, fully connected networks, or hybrid \cite{RN342}. It can be seen from table 5 that USL is the most widely used learning strategies in DL-based dose prediction tasks. A category that falls between USL and SL is SSL. SSL is commonly used for tasks in which the target information is only partially available. GAN, a popular SSL, has also been utilized for dose prediction tasks (Table 5). 

\begin{figure}
\centering
\noindent \includegraphics*[width=6.50in, height=4.20in, keepaspectratio=true]{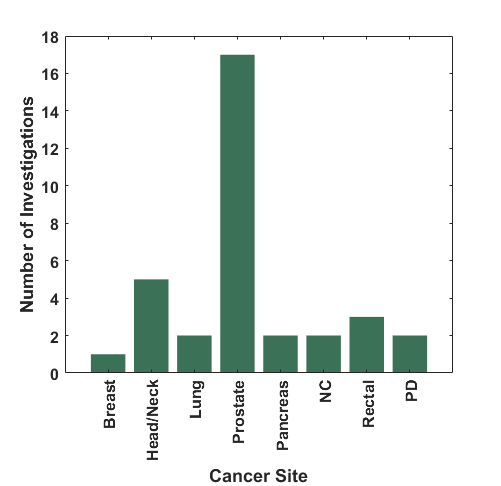}

\noindent NC = Nasopharyngeal Cancer; PD = Personalized Dosimetry \\
Figure 4. The total number of DL-based dose prediction investigations for various cancer sites.

\end{figure}

\noindent \textbf{Table 4. }A list of articles with investigations on effects of outliers on plan quality and summary of evaluation metrics used by RapidPlan${}^{TM}$ with threshold in parentheses.\textbf{}

\begin{longtable}{|p{1.2in}|p{1.6in}|p{1.4in}|} \hline 
\textbf{Reference} & \textbf{Method} & \textbf{Outlier} \\ \hline 
\cite{RN98} & Restricted sum of residual (RSR) & Dosimetric \\ \hline 
\cite{RN284} & Regression and residual analysis & Dosimetric \\ \hline 
\cite{RN136} & Leverage and studentized residual & Dosimetric, Geometric \\ \hline 
\cite{RN325} & Regression analysis scatter plots, cook's distance & Dosimetric,\newline Geometric \\ \hline 
\cite{RN237} & Model-based case filtering & Dosimetric, Geometric \\ \hline 
RapidPlan${}^{TM}$ & Cook's distance ($\mathrm{>}$10)\newline Studentized residual ($\mathrm{>}$ 3)\newline Modified Z-score ($\mathrm{>}$3.5)\newline Areal Difference of Estimate ($\mathrm{>}$3) & Dosimetric\newline and\newline Geometric \\ \hline 
\end{longtable}

3.3.4 Influence of various parameters on DL-based model performance \\

\textbf{Input parameters}

In terms of number of input parameters, Williems \textit{et al.} studied the impact of four different inputs (Table 5) for dose prediction under with and without data normalization of dose distribution. The order of models in terms of performance was CT + isocenter + contours > CT + contours > CT + isocenter > CT only. While the dose distribution normalization had more benefits for CT + contours, it was found to be less necessary for CT + isocenter + contours model. Whereas, normalization produced hot and cold spots for CT + isocenter model \cite{RN1}.
While many studies use only CT with anatomical information (i.e. PTV and OAR contours) as inputs to the CNN \cite{RN7, RN10, RN111} as can be seen in Table 1, Barragán‐Montero \textit{et al.} included beam gemoetry information along with anatomical information as inputs. As a result, the model was able to learn from database that was heterogeneous in terms of beam configurations (i.e. noncoplanar) \cite{RN7}, which was the limitation of network proposed in the earlier studies \cite{RN111}. For rectal cancer IMRT, Zhou \textit{et al.} showed improvements in the prediction accuracy by including beam configurations as input to the network compared to that of without beam configuarations \cite{RN290}. For head/neck cancer, Chen \textit{et al.} investigated the influence of adding out of field labels into the network training to deal with inability of 2D network to account for radiation beam geometry. It resulted in a better overall performance compared to the network excluding out-of-field labels \cite{RN9}. For prostate cancer, Murakami \textit{et al.} compared the performance of CT-only based GAN with contour-based GAN in predicting target images (i.e., RT-dose images) and found prediction performance of contoured-based GAN to be superior. 

\textbf{Loss functions}

In terms of losses, MSE is one of the most widely used cost functions in DL methods as it has many desirable properties from an optimization standpoint. Owing to its simplicity, well behaved gradient and convexity, majority of previous studies including the ones shown in Table 1 utilized only MSE loss for dose prediction. Nguyen \textit{et al.} trained network with domain-specific loss function by adding nonconvex DVH and adversarial loss in addition to MSE loss function. While this outperformed dose predictions compared to MSE based trained model, for the same computational system, it increased the training time to 3.8 days with 100000 iterations compared to 1.5 days for MSE only based network \cite{RN264}. 
Lee \textit{et al.} and Chen \textit{et al.} utilized mean absolute error (MAE) cost function between the ground truth and dose rate map predicted by CNN \cite{RN9, RN267}. A key difference between MSE and MAE that MAE is more robust to outliers but may be inefficient to find the solution, whereas MSE provides more stable and closed form solution. Other loss functions may include Huber loss, smooth mean absolute error, quantile loss, and log cosh loss function. So far, MSE loss function has been the standard cost function used in DL-based dose prediction studies.

\textbf{Sample size }

In general, the DL based methods require a large number of high quality data to be effective. A small datasets in DL can be challengening as it can result in overfitting. Overfitting occurs when the model is trained to exactly fit a set of training data, however cannot learn the hidden pattern to maintain model generality \cite{RN342}. 
Data augmentation \cite{RN4}, dropout layer  \cite{RN312}, estimation based on the training and the validation curves \cite{RN10}, synthesizing new data based on physics principles \cite{RN313} or incorporating regularizations to model parameters \cite{RN314} have been used in the literature to prevent overfitting. The process of data augmentation, more commonly used in dose prediction approaches, is to expand dataset by synthesizing additional realistic samples from available samples. It is important to note here, however, that the process of augmentation to be used depends on the suitability of the context. For the purpose of dose prediction, we have presented the average training and testing sample size for each treatment site in Figure 5 for all DL-based dose prediction methods to date, which provides the readers with an approximate range of training and testing data set for each cancer site. 

As shown in Table 1, three investigations on prostate cancer have been reported so far for predicting pareto optimal dose distirbutions \cite{RN108, RN265, RN264}. For each patient in training set, 10, 100, and 500 plans were generated by Ma \textit{et al.}, Nguyen \textit{et al.}, and Bohara \textit{et al.}, respectively, to sample the pareto surface with different tradeoffs. An optimal number of plans per patient in training set is unknown as it may depend on case to case basis. Nonetheless, in the case of predicting pareto optimal plans, it may be ideal to stay within clinically relevant regime by including only those plans that covers dosimetric tradeoffs presented by a physician.

Kandalan \textit{et al.} studied the issue of generalizing DL-based dose prediction models and to make use of transfer learning to adapt a DL dose prediction model to different planning styles in the same institutions and planning practices at different institutions. A source model was adapted to four different planning styles only with 14-29 cases \cite{RN268}. A long-term goal of these studies is to generate a universal model that can easily be transferred to different institutions for a similar task.

\noindent \textbf{Table 5. }A list of publications on DL-based dose prediction for various treatment sites.\textbf{}

\begin{longtable}{|p{0.9in}|p{0.8in}|p{2.2in}|p{0.5in}|} \hline 
\textbf{Reference} & \textbf{Architecture} & \textbf{Input} & \textbf{Output} \\ \hline 
\cite{RN34} & ANN & Number of fields, PTV volume, PTV to OAR distance, azimuthal and elevation angles  & 3D \\ \hline 
\cite{RN44} & ANN & Distance to PTV, Distance to OARs\newline PTV volume & 3D \\ \hline 
\cite{RN273} & ANN & 16 different geometrical parameters & 3D \\ \hline 
\cite{RN111} & Modified\newline U-net & PTV + OAR + Prescription & 2D  \\ \hline 
\cite{RN3} & ResNet-50 & CT + OAR + PTV images + dose distribution image & 2D  \\ \hline 
\cite{RN5} & 3D-FCN & 3D CT + OAR + Prescription & 3D  \\ \hline 
\cite{RN90} & U-Res-Net & 3D CT + OAR & 3D  \\ \hline 
\cite{RN11} & GAN & Contoured CT images + dose distribution & 2D \\ \hline 
\cite{RN7} & HD U-Net & OAR + PTV + Beam information with approximated dose & 3D  \\ \hline 
\cite{RN9} & CNN -\newline Res-Net 101 & Contoured images + coarse dose map, with out of field labels  & 2D   \\ \hline 
\cite{RN267} & U-Net & PET and CT image patches & 3D  \\ \hline 
\cite{RN10} & HD U-Net & OAR + PTV & 3D  \\ \hline 
\cite{RN1} & U -- Net & CT only,\newline CT + ISO, CT + Contours, CT + ISO + Contours & 2D \\ \hline 
\cite{RN265} & Modified 3D U-Net & DVHs + Contours & Pareto Dose Distributions  \\ \hline 
\cite{RN108} & U-Net & PTV + Body + OAR,\newline PTV + Body + OAR + Dose information from selected beam angles & Pareto Dose Distributions \\ \hline 
\cite{RN269} & 3D U-Net DRN & CT + FMCV & 3D  \\ \hline 
\cite{RN105} & Modified U-Net & Density map + 3D CT +Activity map & 2D  \\ \hline 
\cite{RN110} & U-Net & PTV + OAR contours & 3D  \\ \hline 
\cite{RN6} & GAN & CT + RT Doses,\newline PTV + OAR & 2D \\ \hline 
\cite{RN272} & ResNet-50 & CT + OAR +PTV +body contours & 3D \\ \hline 
\cite{RN4} & U-Net & Low resolution dose + CT & 2D \\ \hline 
\cite{RN107} & HD U-Net & CT + RT dose distribution & 3D \\ \hline 
\cite{RN262} & GAN & CT + PTV + OAR & 2D \\ \hline 
\cite{RN261} & Attention gated GAN & CT + PTV + OAR & 3D  \\ \hline 
\cite{RN264} & GAN & PTV + OAR + Body & Pareto Dose Distribution  \\ \hline 
\cite{RN285} & 3D GAN & Contoured CT images & 3D  \\ \hline 
\cite{RN290} & 3D U-Net + Residual Network & CT + OAR + PTV contours + Beam + Dose & 3D  \\ \hline 
\cite{RN268} & 3D U-Net & OAR + PTV contours & 2D  \\ \hline 
\cite{RN339} & Dense-Res hybrid Network  & Beam + structural information & Static field fluence prediction  \\ \hline 
\cite{RN340} & Virtual Treatment Planner Network  & DVH & TPPs adjustment action  \\ \hline 
\end{longtable}

FMCV = Fluence Map Converted Volume; PTV = Planning Target Volume; OAR = Organ at Risk; GAN = Generative Adversarial Network; HD = Hierarchically Dense; DVH = Dose Volume Histograms; TPP = Treatment Planning Parameter

\begin{figure}
\centering
\noindent \includegraphics*[width=6.50in, height=4.20in, keepaspectratio=true]{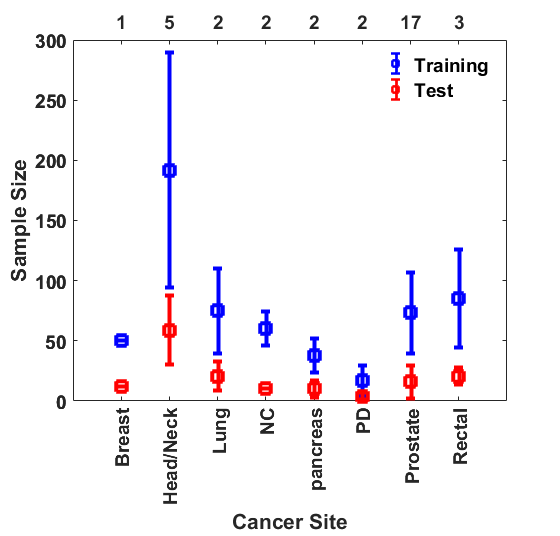}

\noindent Figure 5. The average training and testing sample size in DL-based dose prediction methods for each cancer site. The values are averaged over number of investigations listed on top x-axis and the error bars represent standard deviation.

\end{figure}

\noindent 
\section{Discussion}

With the aim of minimizing the variations in treatment planning and improving the treatment planning efficiency – namely a time-consuming trial-and-error process of planning a treatment from scratch for every patient, the researchers introduced the concept of using previously delivered treatment plans in order to guide treatment planning for a new patient. This concept has been labelled as a knowledge-based planning today. In the last decade, there has been a rapid growth in the number publications in traditional KBP dose prediction. On the other hand, the number of publications on DL has increased exponentially in recent years owing to its flexibility and superior performances compared to many state-of-the-art techniques. Over 90 papers have been published on traditional KBP dose prediction methods between 2011 and August 2020, whereas over 15 publications have already been published on DL-based dose prediction this year so far. 
In general, most paper demonstrated improvements in comparison to manually optimized clinical plans in terms of both treatment planning quality and efficiency. A large number of manuscripts were published on traditional methods between 2015 – 2018, with the highest number of publications in 2017. This is presumably due to commercialization of the RapidPlanTM in Eclipse® treatment planning software in 2014, which allowed researchers from different centers to perform range of retrospective studies for investigating the influence of various parameters on the quality of plans generated through RapidPlanTM KBP. While the number of traditional KBP based publications has been quite steady in the past 2 years, the DL-based publications have been rapidly increasing since 2017. 
In terms of modality, both techniques were mostly applied to IMRT, VMAT and other noncoplanar intensity modulated external beam radiation therapy treatments. Only a small number of data driven dose prediction studies were reported for the purpose of magnetic resonance imaging guided therapy (MRgRT) \cite{RN273}. The number of traditional KBP and DL based publications for on-table adaptation may increase in the future, owing to recent technical developments such as MR-Linear Accelerator (MR-Linac). In terms of treatment sites, prostate, head/neck and lung were amongst the most investigated sites in both traditional KBP and DL-based methods compared to complex abdominal or cranial sites. This was anticipated as both KBP techniques require a large training sample set and these three are commonly treated sites in external beam radiation therapy. Therefore, a large repository of previously treated plans is likely to be available for building dose prediction models for these two sites over other complex sites.   
In KBP, three commonly reported dose prediction metrics in the literature were entire DVH curve (Table 1), one or more dose metrics (Table 2) and voxel-based dose prediction (Table 3). A known limitation of DVH prediction is that DVHs are only predicted for contoured OARs, which may limit the accountability of enhance conformity and hotspots that may occur outside of the region of interest. This was addressed through voxel-based dose prediction approaches in which the models are built to predict individual voxels within the CT image. However, this approach relies heavily on the quality of the plans used to build the model as the inclusion of outliers can compromise the model performance. Even for RapidPlanTM based KBP, several studies indicated the need to investigate the proper identification of outlier plans \cite{RN283, RN233, RN278}. Outlier identification in RapidPlanTM involves statistics and regression plots for each structure, suggesting Cook’s Distance > 10.0, Studentized Residual > 3.0, Areal Difference of Estimate > 3, and Modified Z-score > 3.5 as potential outliers \cite{RN328}. To an extent, this also requires removal of outliers in iterative manner with  either stopping the removal once no significant improvement is observed or identification of the outliers followed by re-planning of all the outliers so that it can be reused in the training cohort \cite{RN325}. The time required to address the issue of outliers may vary from one institute to the other as institutions without standardize techniques can have many dosimetric outliers presumably due to a large variations in treatment planning, which, in turn, can result in a time consuming process of eliminating outliers either through visual inspection or additional statistical analysis. In the literature, limited amount of emphasis has been given on establishing a systematic process for identifying dosimetric and geometric outliers. To our knowledge, currently, there is no well-established workflow for outlier identification and mitigation in terms of   model creation for both KBP techniques. Therefore, a standardized automated method of outlier identification and model creation could further enhance the treatment planning experience \cite{RN237}. In contrast to a previous review that presented a number of training and testing sample size per year \cite{RN281, RN341}, we separated the datasets per cancer sites for traditional KBP and DL-based in Figure 3 and 5, respectively. This would provide readers with a range of training sample size for each cancer site, as required number of training set depends not only on the prediction model but also on the complexity of a treatment site. For instance, the number of cases required to train a model may be more for more complex cases such as head/neck to represent the case population versus a simple case such prostate cancer. Direct comparison of training sample size between the traditional and more recent DL-based KBP was not made as DL-based dose prediction is a relatively new technique with a fewer number of investigations per site compared to traditional KBP methods. 
In contrast to DL, an inherent limitation of traditional methods is that it is unable to process the raw data and extract important features and patterns hidden within. Both, similarity measures in atlas-based methods and input features to model-based methods, require considerable effort to extract valuable features (i.e. overlap volume histogram, OAR distance to the PTV, projections, etc.….) that can process raw data either to identify the best matched case or into a representation from which patterns within the input can be classified through a classifier. In traditional approaches, PCA has been widely used in the literature for feature selection owing to its simplicity. However, a major limitation of PCA is that it learns low dimensional representation of data only with a linear projection. Whereas, DNNs can be used to address this issue and untangle non-linear projections. For instance, an autoencoder is a type of neural network that is consisted of encoder, which encodes the input into a low dimensional latent space, and decoder, which restores the original input from the low dimensional latent space \cite{RN179}. This has been adopted in DL-based dose prediction methods (Table 5) and extension of such unsupervised method is anticipated in the near future to further enhance the dose prediction accuracy. 
In terms of DL-based dose prediction methods, two mostly investigated networks, thus far, included CNN and GAN. From the results so far, it appeared that GANs may be a good choice for dose prediction tasks over conventional CNNs for several reasons. First, GAN have been proven to perform well in lesion detection and data augmentation tasks \cite{RN285, RN323}. In addition, GAN does not rely on pure spatial loss, such as mean square error between dose volumes, which makes it a suitable candidate not only for dose prediction of conventional radiation therapy but also for SBRT in which dose heterogeneity is prevalent. Furthermore, Babier \textit{et al.} found that GAN models did not require significant parameter tuning and architecture modifications during implementations compared to other conventional methods \cite{RN285}. However, in contrast to CNN, one limitation of conventional GAN is that they are difficult to train and requires the number of network parameters to be as low as possible. Future studies are anticipated to account for such shortcomings by proposing extension of networks such as attention-gated GAN \cite{RN261}. 

\noindent 
\subsection{Method comparison of KBP dose predictions}

For head/neck cancer, the difference between the traditional KBP predicted and actual median doses for the parotids ranged from -17.7\% to 15.3\% \cite{RN53}, whereas it ranged between – 7.7 to 13.5\% for DL-based dose prediction \cite{RN9}. With the same level of prediction accuracy, DL-based KBP was able to predict median dose for 80\% of parotids compared to 63\% by the traditional KBP method \cite{RN9}. Kajikawa \textit{et al.} made the direct comparison of dose distribution predicted by DL method with that of generated by RapidPlanTM for prostate cancer \cite{RN110}. This dosimetric comparison showed that CNN significantly predicted DVH accurately for D98 in PTV-2 and V35. V50, V65 in rectum. Given that features automatically extracted by DL methods can include both geometric/anatomic features and the mutual tradeoffs between the OARs, it gives an edge to DL-methods in terms of dose prediction accuracy compared to traditional KBP methods that mainly rely on DVH and geometry-based expected dose.
For oropharyngeal cancer, Mahmood \textit{et al.} directly compared GAN approach for generating predicted dose distribution with several traditional approaches including bagging query \cite{RN72, RN43} and generalized PCA \cite{RN37}, random forest \cite{RN69}. Mahmood \textit{et al.}, through the gamma analysis \cite{RN318}, demonstrated that GAN plans were the most similar to the clinical plans and achieved 4.0 \% to 7.6 \% improvements in frequency of clinical criteria satisfactions compared to traditional approaches \cite{RN11}. 
	For prostate SBRT, Vasant  \textit{et al.} compared the performance of proposed attention gated GAN with an earlier approach that used relative distance map information of neighboring input structures \cite{RN34}. In contrast to conventional radiation therapy, SBRT produces hot spots within the target volume. Mean absolute difference in V120 between KBP like approach and actual plan was four-fold higher compared to that achieved by attention gated GAN technique, demonstrating the ability of a DL-based method to predict cold spots and hotspots that are prevalent to SBRT dose volumes.  
Both, traditional and DL-based, KBP approaches used the data from previously treated patients to make dose prediction for a new patient. DL based approaches, however, have been shown to outperform traditional methods in dose prediction tasks as demonstrated by several studies in the literature. This is presumably due to ability of DL-methods to not be limited by a small number of features in contrast to that of in traditional KBP. 

\noindent 
\subsection{Future trends}

From the statistics of publications on data driven dose prediction approaches in recent years, there is a clear trend of transformation from traditional methods to DL-based methods for KBP. This is presumably due to flexibility and superior performance of DL based approaches in contrast to traditional approaches. In terms of traditional KBP methods, future investigations are anticipated to be retrospective in nature by using clinically available tools (i.e., RapidPlanTM). On the other hand, DL based methods appear to be in its initial development stage, hence, its potential will be explored in different areas of dose prediction tasks in treatment planning workflow including adaptive radiotherapy in near future. 
Adaptive radiotherapy (ART) involves adjusting dose distribution based on anatomical changes observed on intra-procedural imaging such CBCT. The standard approach requires physician to perform recontouring of OAR and tumor regions followed by plan re-optimization, which is difficult to implement in an ART. To date, only one study has been reported to adopt DL methods for the purpose of ART of head/neck cancer \cite{RN262}. The future trend will certainly be towards utilizing the flexibility and efficiency offered by DL-based methods to present dose prediction models of dosimetry changes and radiotherapy response for ART.  
Post dose prediction, a main component of treatment planning workflow includes ensuring the achievability of the predicted dose plans, which often involves inverse treatment planning through manual intervention. Only handful studies extended such data driven approaches in a fully automated pipeline that not only predict the dose distribution but also generates a complete treatment plan with minimum human interaction in traditional \cite{RN332, RN316, RN244, RN151, RN35} and DL based methods \cite{RN285, RN339, RN11}. The deliverability of the predicted plans is more important as it has to account for various mechanical and algorithmic constraints. It is important to note here that good predictions with low error may not necessarily lead to the final deliverable plan with the same performance on clinical criteria. For instance, five of the seven prediction methods investigated by Babier \textit{et al.} resulted in a significantly worse clinical criteria satisfaction despite lower error post dose predictions \cite{RN285}. We, therefore, believe synchronizing an inverse optimization engine with dose prediction methods hold a great potential in improving treatment planning efficacy and efficiency. Alternatively, a DL-based fluence prediction has also been proposed for real-time prostate treatment planning \cite{RN339}. This approach follows conversion of predicted fluence maps to a deliverable treatment plan through delivery parameter generation and dose calculations directly in a treatment planning software. Such approaches do not require inverse optimization process and involve minimal human intervention. A subsequent task, after generating a deliverable plan, involves patient specific quality assurance measurements that are performed routinely prior to actual treatment delivery to ensure delivery and dosimetry accuracies. Several ML \cite{RN319} and DL \cite{RN321} approaches have been reported for predicting gamma passing rates for IMRT patient specific QA. More efforts are also anticipated to be placed to incorporate such approaches into treatment planning pipeline to establish a fully automated workflow. 
One of the challenges in data driven algorithms, including both ML and DL, is that it requires a large set of a high-quality data. Since the quality of data and radiotherapy practices vary from one center to the other, the heterogeneity in previously treated plans become a major obstacle in deployment of data-driven solutions in the field of radiation oncology. To address this issue, the concept of transfer learning for model adaptation to different learning styles at different centers may be investigated further in the future. A long-term goal of this area of investigations would be to incorporate data-driven predictive tools as a part of the clinical pathway.

\noindent 
\section{Conclusion}

In the last decade, a tremendous amount of work has been done towards automation to improve treatment planning quality and efficiency. We have performed a review of two major KBP approaches to dose prediction: traditional KBP methods with over 90 articles and more recently introduced DL-based KBP with nearly 30 articles. While traditional approaches are either equivalent or superior to an experienced planner with greater efficiency, recent developments in DL holds a greater potential in dose prediction task. Both KBP approaches, however, are needed to be expanded for more complex sites such as abdominal and intercranial. Given commercial accessibility of RapidPlanTM module, more retrospectives studies are foreseen in the future. However, new approaches DL-based KBP are actively being introduced and trending in a steep upward direction. There are various areas of future research, several of which have been highlighted in this review, required to achieve an ultimate goal of a fully automated treatment planning system. 

\noindent 
\bigbreak
{\bf Acknowledgements}

This research is supported in part by the National Cancer Institute of the National Institutes of Health under Award Number R01CA215718, the Department of Defense (DoD) Prostate Cancer Research Program (PCRP) Award W81XWH-17-1-0438 and Dunwoody Golf Club Prostate Cancer Research Award, a philanthropic award provided by the Winship Cancer Institute of Emory University.

\noindent 
\bigbreak
{\bf Disclosures}

The authors declare no conflicts of interest.

\noindent 

\bibliographystyle{plainnat}  
\bibliography{Manuscript_tex2}      

\end{document}